\documentclass[prl,twocolumn]{revtex4}

\usepackage{graphicx}
\usepackage{amsmath}
\usepackage{epsfig}

\begin{document}

\title{Wide-bandwidth, tunable, multiple-pulse-width optical delays
using slow light in cesium vapor}

\author{Ryan M. Camacho, Michael V. Pack, John C. Howell}
\affiliation{Department of Physics and Astronomy, University of
Rochester, Rochester, NY 14627, USA }
\author{Aaron Schweinsberg, Robert W. Boyd}
\affiliation{Institute of Optics, University of Rochester,
Rochester, NY 14627, USA }

\begin{abstract}
We demonstrate an all-optical delay line in hot cesium vapor that
tunably delays 275 ps input pulses up to 6.8 ns and 740 input ps
pulses up to 59 ns (group index of approximately 200) with little
pulse distortion. The delay is made tunable with a fast
reconfiguration time (100's of ns) by optically pumping out of the
atomic ground states.
\end{abstract}

\maketitle \

There is considerable practical interest in developing all-optical
delay lines that can tunably delay short pulses by much longer than
the pulse duration. Slow light (i.e.\ the passage of light pulses
through media with a small group velocity) has long been considered
a possible mechanism for constructing such a delay line. Most
commonly, the steep linear dispersion associated with a single gain
or transparency resonance provides the group delay. Most early work
used the dispersion associated with electromagnetically induced
transparency \cite{kasapi95,kash99,Budker99,hau99,phillips01,liu01,
turukhin02}, but recently other resonances have been explored,
including coherent population oscillations
\cite{bigelow03,Zhao05,palinginis05}, stimulated Brillouin
scattering
\cite{song05a,Okawachi05,gonzalez-herraez05,gonzalez-herraez06},
stimulated Raman scattering \cite{sharping05,dahan05}, and spectral
hole-burning \cite{camacho06_2}.

In addition to single-resonance systems, double gain resonances have
been used for pulse advancement
\cite{chiao93,wang00,dogariu01,stenner03,macke03,agarwal04} and
delay \cite{stenner05}.  Widely spaced gain peaks create a region of
anomalous dispersion, resulting in pulse advancement. When the
spacing between the gain peaks is small, a region of normal
dispersion is created, resulting in pulse delay.  Pulse advancement
is also possible by the proper spacing of two absorbing resonances
\cite{macke03}.  The possibility of pulse delay between two
absorbing resonances has also received some attention
\cite{grischkowsky73,tanaka03,macke06,camacho06,zhu06,camacho07_1}.

Ideally, an optical delay line would delay high bandwidth pulses by
many pulse lengths in a short propagation distance without
introducing appreciable pulse distortion and be able to tune the
delay continuously with a fast reconfiguration rate. Minimal pulse
absorption is also desirable, but not necessary because absorption
can be compensated through amplification. Relatively few experiments
\cite{kasapi95,hau99,turukhin02,song05a,camacho06,camacho06_2} have
directly measured pulse delays longer than the incident pulse
duration, and of these, none has used pulses shorter than 2 ns or
reported reconfiguration rates approaching the inverse pulse delay
time.

In this Letter, we demonstrate the tunable delay of a
1.6-GHz-bandwidth pulse by up to 25 pulse widths and the tunable
delay of a 600-MHz-bandwidth pulse by up to 80 pulse widths by
making use of a double absorption resonance in cesium.  Furthermore,
we show that the delay can be tuned with a reconfiguration time of
100's of nanoseconds.

\begin{figure}[t]
\centerline{\includegraphics[scale=0.4]{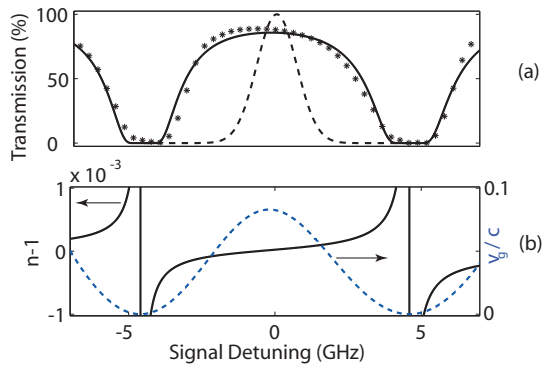}}
\caption{(a) CW signal transmission (asterisks--measured, solid--fit
) overlayed with the spectrum (dashed) of a 275 ps pulse and (b)
index of refraction (solid) and group velocity (dashed), all versus
signal detuning for cesium at approximately 114 $^{\circ}$C.  All
theory curves taken from  Eq.\ (\ref{eq:index}) with $\mathcal{A} =
4 \times 10^5$ rad/s, $g_1 = 7/16$ and $g_2 = 9/16$. High-fidelity
optical delay is observed for light pulses passing through the
nearly transparent window between the two resonances. }
\label{resonance}
\end{figure}

In a medium with two Lorentzian absorption resonances, as
illustrated in Figure \ref{resonance}, the complex index of
refraction can be approximated as
\begin{equation}\label{eq:index}
n(\delta) = 1 - \frac{\mathcal{A}}{2} \left( \frac{g_1}{\delta +
\Delta_+ + i\gamma} + \frac{g_2}{\delta - \Delta_- + i\gamma}
\right)
\end{equation}
where $2\gamma$ is the homogeneous linewidth (full width at half
maximum, FWHM), $g_1$ and $g_2$ account for the possibility of
different strengths for the two resonances,
$\delta=\omega-\omega_0-\Delta$ is the detuning from peak
transmission, $\omega_0=(\omega_1+\omega_2)/2$, $\omega_1$
($\omega_2$) is the resonance frequency for transition 1 (transition
2), $\Delta_{\pm}=\omega_{21}\pm\Delta$,
$\omega_{21}=(\omega_{2}-\omega_{1})/2$, and
\begin{equation}
\Delta = \frac{g_1^{1/3}-g_2^{1/3}}{g_1^{1/3}+g_2^{1/3}}\omega_{21}.
\end{equation}
For example, alkali atoms have two hyperfine levels associated with
their electronic ground state, leading to two closely spaced
absorption resonances.  We note that any other system with two
similar absorbing resonances may also be used (e.g. quantum dots,
microresonators, photonic crystals, etc).  For a vapor of alkali
atoms, the detunings satisfy $\Delta_+ \approx \Delta_-\gg\gamma$,
and the strength of the resonance is given in SI units by
$\mathcal{A} = N|\mu|^2 / [\epsilon_0\hbar (g_1+g_2)]$, where $\mu$
is the effective far-detuned dipole moment \cite{steck03}, and $g_1$
and $g_2$ are proportional to the degeneracies of the hyperfine
levels.

Eq.\ (\ref{eq:index}) is also applicable for inhomogeneously
broadened lines, such as Doppler broadened atomic vapors, if the
detunings $\Delta_-$ and $\Delta_+$ are greater than the
inhomogeneous linewidth by an order of magnitude or more. This
result holds because the homogenous Lorentzian lineshape has long
wings while the inhomogeneous lineshape decreases exponentially.

By expanding Eq.\ (\ref{eq:index}) about the point $\delta=0$, we
find that the real part $n^{\prime}$ and imaginary part
$n^{\prime\prime}$ of the index of refraction are given by
\begin{subequations}\label{eq:n_real_imag}
\begin{eqnarray}\label{eq:n_real}
n^{\prime}(\delta)&\approx& 1+K_0+ K_1\frac{\mathcal{A}
}{\omega_{21}^2}\delta+ K_3\frac{\mathcal{A}
}{\omega_{21}^4}\delta^3
\end{eqnarray}
\vspace{-0.5cm}
\begin{eqnarray}\label{eq:n_imag}
n^{\prime\prime}(\delta)&\approx& K_1\frac{\mathcal{A}
\gamma}{\omega_{21}^2}+3K_3\frac{\mathcal{A}
\gamma}{\omega_{21}^4}\delta^2,
\end{eqnarray}
\end{subequations}
where
\begin{equation}
K_i=
\left(\frac{g_1^{1/3}+g_2^{1/3}}{2}\right)^{i+1}\left(g_1^{(2-i)/3}
+(-1)^{i+1}g_2^{(2-i)/3}\right),
\end{equation}
and where we have assumed that $n-1 \ll 1$ in keeping only the first
few terms in the expansion.  Note that for the special case in which
the two resonances are of equal strength (i.e. $g_1 = g_2=g$), the
coefficients are given by $K_i=2g$ for $i$ odd and $K_i=0$ for $i$
even. For cesium, which has $g_1=7/16$ and $g_2=9/16$, the error
introduced by assuming $g_1 = g_2$ is approximately 0.5\%. For this
reason, we make the simplifying assumption $g_1 = g_2=1/2$
throughout the remainder of the paper.

Pulse propagation can be described in terms of various orders of
dispersion, which can be determined through use of Eq.\
(\ref{eq:n_real}) as
\begin{equation}\label{eq:disp}
\beta_j=\left.\frac1c\frac{d^j\omega
n'(\omega)}{d\omega^j}\right|_{\omega=\omega_0+\Delta},
\end{equation}
Thus the group velocity is given by $v_g=1/\beta_1$, and the
group-velocity dispersion (GVD) and third-order-dispersion are given
respectively by $\beta_2$ and $\beta_3$. The absence of second-order
(first-order) frequency dependence in Eq.\ (\ref{eq:n_real}) (Eq.\
(\ref{eq:n_imag})) means that near $\delta=0$ the GVD (absorption)
is minimized regardless of possible differences between $g_1$ and
$g_2$. Thus, between two absorption resonances, which can be
described by Eq.\ (\ref{eq:index}), the maximum transparency is
accompanied by a minimum in GVD.

We next develop a simple model to provide an understanding of the
role of dispersion and absorption on pulse broadening.  We
provisionally define the pulse width as the square root of the
variance of the temporal pulse shape. For an unchirped Gaussian
pulse, i.e. $E(0,t)=E_0\exp\left(-t^2/2T_0^2\right)$, the pulse
width defined in this way is simply $T_0$. The pulse width after
propagating through a distance $L$ of dispersive medium is then
given to third-order in $\delta$ by \cite{agrawal95}
\begin{eqnarray}\label{eq:agarwal}
T_d^2 &=& T_0^2 + \left(\frac{2\beta_2 L}{T_0} \right)^2 +
\left(\frac{\beta_3 L}{2T_0^2} \right)^2
\end{eqnarray}
where $T_0$ is the initial pulse width. In the case of cesium, where
$\omega_0 \approx 2\pi \times 3.5 \times 10^{14} $ rad/s and
$\omega_{21} \approx \pi \times 9.2 \times 10^9 $ rad/s, $\beta_2$
can be neglected and Eq.\ (\ref{eq:agarwal}) simplifies to
\begin{eqnarray}\label{eq:dispersive_broadening}
T_d^2 &\approx& T_0^2+\left(\frac{3\tau_d}{\omega_{21}^2
T_0^2}\right)^2,
\end{eqnarray}
where $\beta_3$ has been calculated using Eqs.\ (\ref{eq:disp}) and
(\ref{eq:n_real_imag}) and where $\tau_d \approx \alpha_0 L/2\gamma$
is the pulse delay and $\alpha_0 L = 2 \omega_0 n''L/c$ is the
optical depth at the pulse carrier frequency $\omega_0$. We further
note that the change in pulse width due to absorption only can be
approximated as \cite{boyd05,camacho06}
\begin{eqnarray}\label{eq:absorptive_broadening}
T_a^2 &=& T_0^2 + \frac{6 \gamma \tau_d}{ \omega_{21}^2 },
\end{eqnarray}
so long as $(T_a/T_0-1)<1$.  

The fractional broadening due to dispersion, defined as $T_d/T_0-1$,
scales as $1/T_0^3$, while the broadening due to absorption scales
as $1/ T_0$. In the present study, $\tau_d\approx 10^{-8} $ s,
$\omega_{21} \approx 10^{11} $ rad/s, $T_0 \approx 10^{-10} $ s, and
$\gamma \approx 10^7 $ rad/s, indicating that dispersion is the
dominant form of broadening by about three orders of magnitude, and
the absorptive contribution to broadening can be ignored.

Experimentally it is much easier to quantify  pulse widths in terms
of their FWHM rather than in terms of their variance as we have done
in Eqs.\ (\ref{eq:agarwal}) - (\ref{eq:absorptive_broadening}). In
the remainder of this Letter, we will quote pulse widths in terms of
their FWHM.

\begin{figure}[t]
\centerline{\includegraphics[scale=0.55]{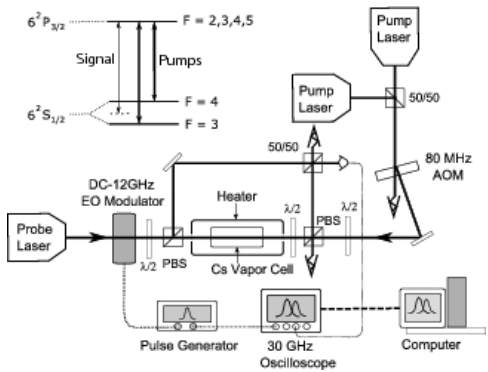}}
\caption{ Experimental schematic.  A signal pulse passes through a
heated cesium vapor cell.  Two pump beams combine on a beamsplitter
and counter-propagate relative to the signal beam through the vapor,
to provide tunable delay of the signal pulse. } \label{experiment}
\end{figure}

\begin{figure}[t]
\centerline{\includegraphics[scale=0.4]{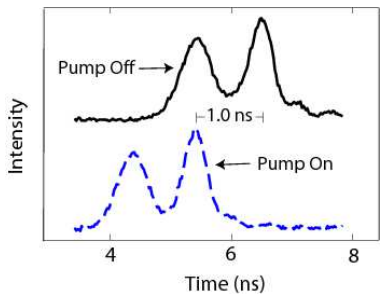}}
\caption{Pulse output waveforms with auxiliary pump beams on
(dotted) and off (solid).  Two 275 ps input pulses separated by 1 ns
are delayed by approximately 5.3 ns without pumping, but only 4.3 ns
with pumping (a change of one bit slot) with little change in pulse
shape. } \label{pulsetrain}
\end{figure}

\begin{figure}[t]
\centerline{\includegraphics[scale=0.36]{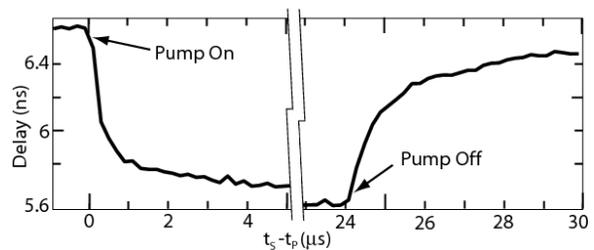}}
\caption{Pulse delay versus time following pump turn-on and
turn-off, showing the reconfiguration time for optically tuning the
pulse delay. The two pump beams are tuned to separate cesium
hyperfine resonances and are switched on at the time origin and
switched off 24 $\mu$s later. } \label{switching}
\end{figure}

Our experimental setup is shown in Fig.\ \ref{experiment}. The
signal laser is a CW  diode laser with a wavelength of 852~nm. The
signal frequency is tuned to obtain maximum transmission between the
two Cs D$_2$ hyperfine resonances and is pulsed at a pulse
repetition frequency of 100~kHz using a fast electro-optic modulator
(EOM). The signal beam is collimated to a diameter of 3~mm, and two
different pulse widths are used, 275~ps or 740~ps FWHM, with a peak
intensity of less than 10~mW/cm${}^2$. The pulses then pass through
a heated 10-cm-long glass cell containing atomic cesium vapor. The
275~ps pulses are measured using a 7.5 GHz silicon photodiode, and
the 740~ps pulses are measured with a 1~GHz avalanche photodiode.
All electrical signals are recorded with a 30 GHz sampling
oscilloscope triggered by the pulse generator. The pump beams are
turned off except for the experiments reported in Figs.\
\ref{pulsetrain} and \ref{switching}.

Figure \ref{resonance}(a) shows the transmission of a CW optical
beam as a function of frequency near the two cesium hyperfine
resonances, overlayed with  the spectrum of a 275 ps Gaussian pulse.
The data points are measured values and the solid line fits these
points to the imaginary part of Eq.\ (\ref{eq:index}). The entire
pulse spectrum lies well within the relatively flat transmission
window between the resonances, resulting in little pulse distortion
from absorption. Figure \ref{resonance}(b) shows the index of
refraction (real part of Eq.\ (\ref{eq:index})) and
frequency-dependent group velocity associated with the absorption
shown in Fig.\ \ref{resonance}(a). We note that, in the region of
the pulse spectrum, the curvature of the frequency-dependent group
velocity is greater than that of the absorption, suggesting that
dispersion is the dominant form of pulse distortion. This is not the
case for single-Lorentzian systems, where the spectral variation of
absorption is the dominant form of distortion \cite{boyd02}. While
most slow light experiments have worked by making highly dispersive
regions transparent, we have worked where a highly transparent
region is dispersive.

\begin{figure}[t]
\centerline{\includegraphics[scale=0.4]{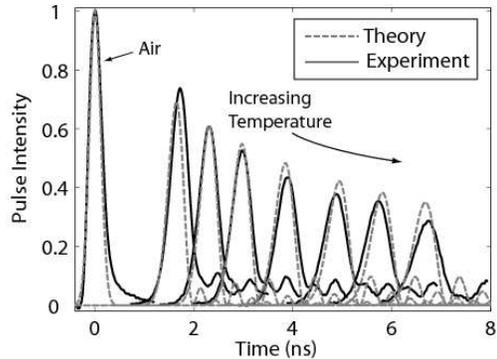}}
\caption{Pulse shapes of 275 ps input pulses transmitted through a
cesium vapor cell. Delays a large as 25 pulse widths are observed.
The temperature range from 90 $^\circ$C to 120 $^\circ$C }
\label{275delay}
\end{figure}

As shown above, the delay of a pulse is proportional to the optical
depth of the vapor.  Figure \ref{275delay} shows that we can control
the delay by changing the temperature (and thus optical depth) of
the Cs cell.  Using a 10 cm cell, and varying the temperature
between approximately 90 $^\circ$C and 120 $^\circ$C, we were able
to tune the delay of a 275 ps pulse between 1.8 ns and 6.8 ns. The
theory curves in Fig.\ \ref{275delay} were obtained using $I(x,t) =
n'(0)c\epsilon_0|E(z,t)|^2/2$ where the electric field is given by
\begin{eqnarray}\label{eq:pulse}
\nonumber E(z,t) =
\frac{E_0T_0\exp\left[-i(\omega_0+\Delta)t\right]}{\sqrt{2\pi}}
\times \\
\int_{-\infty}^\infty d\delta \mbox{exp}\left[i\left(\frac{\omega
n(\delta)}{c}z -\delta t \right)-\frac{\delta^2T_0^2}{2}\right],
\end{eqnarray}
and where we have used Eq.\ (\ref{eq:index}) for the index of
refraction. The atomic density $N$ has been chosen separately to fit
each measured pulse. We note that a pulse may be delayed by many
pulse widths relative to free-space propagation with little
broadening.

\begin{figure}[b]
\centerline{\includegraphics[scale=0.55]{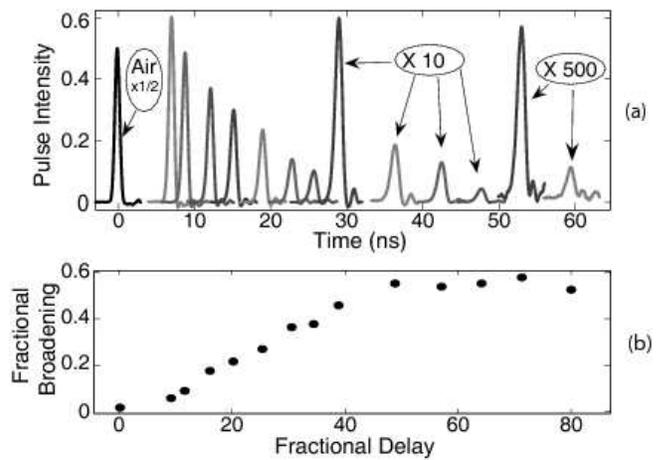}}
\caption{(a) Output pulse shapes and (b) fractional broadening as
functions of fractional delay for a 740 ps input pulse. Fractional
delay is defined as ($\tau_d/T_0$) and fractional broadening is
defined as $(T-T_0)/T_0$. } \label{740ps}
\end{figure}

Longer pulses lead to delay with reduced pulse broadening because
pulse broadening is approximately proportional to $1/T_0^3$ (see
Eq.\ (\ref{eq:dispersive_broadening})). To study the larger
fractional delays enabled by this effect, we used longer 740 ps
input pulses for which the dispersive broadening is significantly
reduced. Figures \ref{740ps}(a) and \ref{740ps}(b) show the delay
and broadening of a 740 ps pulse after passing through a sequence of
three 10 cm cesium vapor cells. The plots correspond to a
temperature range of approximately 110 $^\circ$C to 160 $^\circ$C.
Even though the pulse experiences strong absorption at large delays,
the fractional broadening of the pulse FWHM remains relatively low.

In addition to temperature tuning, the optical depth can be changed
much more rapidly by optically pumping the atoms into the excited
state using two pump lasers. As shown in Fig.\ \ref{experiment} each
pump laser is resonant with one of the $D_2$ transitions in order to
saturate the atoms without optical pumping from one hyperfine level
to the other. The power of each pump beam is approximately 30~mW,
and both pump beams are focused  at the cell center. The signal beam
overlaps the pump beams and is also focused to a 100 $\mu$m beam
diameter. The pump beams are turned on and off using an 80 MHz AOM
with a 100~ns rise/fall time. Being on resonance with the $D_2$
transitions, the pump fields experience significant absorption
($\alpha L\sim300$), and are entirely absorbed despite having
intensities well above the saturation intensity.

With the pump beams on, the decreases in effective ground-state
atomic density leads to smaller delay. Figure \ref{pulsetrain} shows
a delayed pulse waveform consisting of two 275 ps input pulses
separated by 1 ns, with the pump on and off. We note that pump
fields create no noticeable change in the waveform shape or
amplitude. Also, we measured that the change in delay is essentially
proportional to the pump power.

In Fig.\ \ref{switching} the measured signal delay is shown as a
function of the difference between arrival time $t_s$ of the signal
at the cell and the turn-on time $t_p$ of the pump. The rise and
fall times lie in the range 300-600~ns  and vary slightly depending
on the relative detunings of the pumps.

In summary, we have observed large tunable fractional time delays of
high-bandwidth pulses with fast reconfigurations rates and low
distortion by tuning the laser frequency between the two
ground-state hyperfine resonances of a hot atomic cesium vapor cell.
We have shown that in such a medium dispersion is the dominant form
of broadening, and we have characterized the delay, broadening, and
reconfiguration rates of the delayed pulses.

This work was supported by the DARPA/DSO Slow Light program, the
National Science Foundation, andthe Research Corporation.

\bibliographystyle{slowbst}
\bibliography{slowbib}

\begin{thebibliography}{10}
\providecommand{\url}[1]{\texttt{#1}}
\providecommand{\urlprefix}{URL }
\providecommand{\eprint}[2][]{\url{#2}}

\bibitem{kasapi95}
A.~Kasapi \emph{et~al.}, Phys. Rev. Lett. \textbf{74}, 2447 (1995).

\bibitem{kash99}
M.~M. Kash \emph{et~al.}, Phys. Rev. Lett. \textbf{82}, 5229 (1999).

\bibitem{Budker99}
D.~Budker \emph{et~al.}, Phys. Rev. Lett. \textbf{83}, 1767 (1999).

\bibitem{hau99}
L.~V. Hau \emph{et~al.}, Nature \textbf{397}, 594 (1999).

\bibitem{phillips01}
D.~F. Phillips \emph{et~al.}, Phys. Rev. Lett. \textbf{86}, 783 (2001).

\bibitem{liu01}
C.~Liu \emph{et~al.}, Nature \textbf{409}, 490 (2001).

\bibitem{turukhin02}
A.~Turukhin \emph{et~al.}, Phys. Rev. Lett. \textbf{88}, 023602 (2002).

\bibitem{bigelow03}
M.~S. Bigelow \emph{et~al.}, Science \textbf{301}, 200 (2003).

\bibitem{Zhao05}
X.~Zhao \emph{et~al.}, Optics Express \textbf{93}, 7899 (2005).

\bibitem{palinginis05}
P.~Palinginis \emph{et~al.}, Optics Express \textbf{13}, 9909 (2005).

\bibitem{song05a}
K.~Y. Song \emph{et~al.}, Opt. Lett. \textbf{30}, 1782 (2005).

\bibitem{Okawachi05}
Y.~Okawachi \emph{et~al.}, Phys. Rev. Lett. \textbf{94}, 153902 (2005).

\bibitem{gonzalez-herraez05}
M.~Gonzalez-Herraez \emph{et~al.}, Appl. Phys. Lett. \textbf{87}, 081113
  (2005).

\bibitem{gonzalez-herraez06}
M.~Gonzalez-Herraez \emph{et~al.}, Optics Express \textbf{14}, 1400 (2006).

\bibitem{sharping05}
J.~E. Sharping \emph{et~al.}, Optics Express \textbf{13}, 6092 (2005).

\bibitem{dahan05}
D.~Dahan \emph{et~al.}, Optics Express \textbf{13}, 6234 (2005).

\bibitem{camacho06_2}
R.~M. Camacho \emph{et~al.}, Phys. Rev. A \textbf{74}, 033801 (2006).

\bibitem{chiao93}
R.~Y. Chiao, Phys. Rev. A \textbf{48}, R34 (1993).

\bibitem{wang00}
L.~J. Wang \emph{et~al.}, Nature \textbf{406}, 277 (2000).

\bibitem{dogariu01}
A.~Dogariu \emph{et~al.}, Phys. Rev. A \textbf{63}, 053806 (2001).

\bibitem{stenner03}
M.~D. Stenner \emph{et~al.}, Nature \textbf{425}, 695 (2003).

\bibitem{macke03}
B.~Macke \emph{et~al.}, Eur. Phys. J. D \textbf{23}, 125 (2003).

\bibitem{agarwal04}
G.~S. Agarwal \emph{et~al.}, Phys. Rev. A \textbf{70}, 023802 (2004).

\bibitem{stenner05}
M.~D. Stenner \emph{et~al.}, Optics Express \textbf{13}, 9995 (2005).

\bibitem{grischkowsky73}
D.~Grischkowsky, Phys. Rev. A \textbf{7}, 2096 (1973).

\bibitem{tanaka03}
H.~Tanaka \emph{et~al.}, Phys. Rev. A \textbf{68}, 053801 (2003).

\bibitem{macke06}
B.~Macke \emph{et~al.}, Phys. Rev. A \textbf{73}, 043802 (2006).

\bibitem{camacho06}
R.~M. Camacho \emph{et~al.}, Phys. Rev. A \textbf{73}, 063812 (2006).

\bibitem{zhu06}
Z.~Zhu \emph{et~al.}, Opt. Exp. \textbf{14}, 7238 (2006).

\bibitem{camacho07_1}
R.~M. Camacho \emph{et~al.}, Physical Review Letters \textbf{98}, 043902
  (2007).

\bibitem{steck03}
D.~A. Steck. \emph{Cesium {D} line data}, Technical Report No.~LA-UR-03-7943,
  Los Alamos National Laboratory (2003), \texttt{http://steck.us/alkalidata/}.

\bibitem{agrawal95}
G.~P. Agrawal, \emph{Nonlinear Fiber Optics} (Academic Press, 1995), p.~79.

\bibitem{boyd05}
R.~W. Boyd \emph{et~al.}, Phys. Rev. A \textbf{71}, 023801 (2005).

\bibitem{boyd02}
R.~W. Boyd \emph{et~al.}, \emph{Progress in Optics} (Elsevier, 2002), p. 497.

\end{thebibliography}

\end{document}